\documentclass{article}
\pdfoutput=1
\usepackage{lineno,hyperref}
\usepackage{graphicx}
\usepackage{caption}
\usepackage{subcaption}

\usepackage{setspace}
\graphicspath{{fig/}}

\newcommand{\recognizer}{\emph{Recognizer}}
\newcommand{\logic}{\emph{Logic}}
\newcommand{\navigator}{\emph{Navigator}}
\newcommand{\zebra}{\emph{ZebraX}}

\title{Sonification of guidance data during road crossing for people with visual impairments or blindness}

\author{Sergio Mascetti\\
\small{Universit\`{a} degli Studi di Milano}\\
\and
Lorenzo Picinali\\
\small{Imperial College London}\\
\and
Andrea Gerino\\
\small{Universit\`{a} degli Studi di Milano}\\
\and
Dragan Ahmetovic\\
\small{Universit\`{a} degli Studi di Milano}\\
\and
Cristian Bernareggi\\
\small{Universit\`{a} degli Studi di Milano}\\
}

\begin{document}
\maketitle



\begin{abstract}
In the last years several solutions were proposed to support people with visual impairments or blindness during road crossing. These solutions focus on computer vision techniques for recognizing pedestrian crosswalks and computing their relative position from the user.
Instead, this contribution addresses a different problem; the design of an auditory interface that can effectively guide the user during road crossing. Two original auditory guiding modes based on data sonification are presented and compared with a guiding mode based on speech messages.

Experimental evaluation shows that there is no guiding mode that is best suited for all test subjects. The average time to align and cross is not significantly different among the three guiding modes, and test subjects distribute their preferences for the best guiding mode almost uniformly among the three solutions.
From the experiments it also emerges that higher effort is necessary for decoding the sonified instructions if compared to the speech instructions, and that test subjects require frequent `hints' (in the form of speech messages).
Despite this, more than $2/3$ of test subjects prefer one of the two guiding modes based on sonification. There are two main reasons for this: firstly, with speech messages it is harder to hear the sound of the environment, and secondly sonified messages convey information about the ``quantity'' of the expected movement.
\end{abstract}





\section{Introduction}
Mobile devices provide new exciting opportunities for people with Visual Impairments or Blindness (VIB).
Indeed, most commercial devices (e.g., based on iOS and Android) are accessible to people with VIB\footnote{In case the reader is unfamiliar with accessibility tools for visually impaired individuals, we suggest a  short video introducing the main ideas - \url{http://goo.gl/mEI6Uz}}.
On one hand, this allows people with VIB to use most of the applications available on mobile devices, such as web browsers and email clients. On the other hand, accessible mobile devices can be used to implement assistive technologies, with great advantages for both developers and users.
The developers can rely on well known platforms, for which there is plenty of documentation and software libraries, and which provide high level OS APIs to support accessibility (e.g., text-to-speech functionalities on iOS).
For the final user, a single device capable of providing different assistive tools is cheaper, quicker to learn and more convenient (in terms of weight to carry, devices to charge, etc...).

Mobile devices also have two main advantages with respect to traditional ones (i.e., desktops and laptops).
Firstly, they can be used on the move, hence can provide support in many situations in which it is impractical to rely on a traditional device.
Secondly, mobile devices are equipped with hardware sensors such as GPS receivers, accelerometers, and gyroscopes, that can be used to acquire information about the user's context and position.
In this context, it is not surprising that several research contributions in the last years focused on mobile assistive technologies.
In particular, a number of solutions have been proposed to support autonomous mobility, for example by recognizing objects in the environment and notifying the user accordingly.

In this contribution we take into account the problem of guiding the user towards and over a zebra crossing (i.e., a particular type of pedestrian crosswalk also called ``continental crosswalk'' in the United States).
This problem involves non-trivial computer vision techniques to recognize the zebra crossing pattern, as well as advanced spatial reasoning, based also on accelerometer data, to reconstruct the position of the crosswalk with respect to the user.
In our previous work we describe the `recognition' procedure used to identify the crosswalk and compute its relative position \cite{AhmetovicICPR2014}.

Other existing contributions in the field focus on the recognition procedure \cite{se,uddin1,uddin2,ivanchenko,ivanchenko2,dragan}. However, a different challenge is now arising: how to guide the user employing audio instructions.
Two contrasting objectives emerge. On one hand audio instructions should provide precise and responsive information. On the other hand, they should not distract the user's attention from the surrounding environment.

This paper presents two auditory guiding modes based on data sonification.
The two guiding modes are similar, with the main difference being that one produces mono sound (i.e., one single sound signal) and the other produces stereo sound (i.e., two different sound signals, one for the left and one for the right ear).
From the applicative point of view, a major difference can be noted; stereo sonification requires the user to wear headphones, while mono sonification can also be reproduced from the device's internal speaker.

The sound design process was conducted employing a user-centric approach, frequently considering end users feedback and carrying out a preliminary evaluation session. The two sonifications, together with a guiding mode based on speech messages, have been implemented in the \zebra{} prototype, an iPhone application that adopts a state-of-the-art algorithm to detect zebra crossings.
\zebra{} was then used to conduct three sets of evaluations aimed at assessing the effectiveness of the guiding modes.
Experimental results show that the three guiding modes can effectively support the test subject to align with the zebra crossing and to actually cross it.
Still, the two guiding modes based on sonification are less immediate to use, and some subjects required frequent hints (in the form of speech messages) to correctly interpret the sonified instructions.

Despite this, two results are available supporting the applicability of the two guiding modes based on sonification.
Firstly, after a few minutes of training only, there is not a statistically significant difference in the performance (e.g., crossing time) between the three guiding modes.
Secondly, $75\%$  of  the  subjects  declared  that  they preferred  the two guiding modes based on sonification. Furthermore, they reported that hearing sounds from the surrounding environment, a very important task when crossing a road, is more difficult with the speech mode than with the two sonifications.

Section~\ref{sec:background} describes the related work as well as the system architecture of \zebra{}. The three auditory guiding modes are presented in Section~\ref{sec:guidingModes} while Sections~\ref{sec:intEval} and \ref{sec:finalEval} present the results of two evaluation sessions. Section~\ref{sec:concl} concludes the paper and highlights future work.


\section{Background}
\label{sec:background}
It is well known that independent mobility is very challenging for people with VIB.
Blind people can find their way by means of a white cane or a guide dog, whereas partially sighted people can also rely on their residual sight.
The main difficulties are related with avoiding obstacles along the way (e.g., people on the sidewalk, trash bins, poles, etc.), finding a target (e.g., stairs, doors, intersections, etc.) and getting information reported on pedestrian signs (e.g., crossing a road over a zebra crossing when the traffic light is green, etc.).

Over the years, many solutions for supporting independent mobility have been investigated in scientific literature. In particular, in the following paragraphs we report the main findings in the field of pedestrian crosswalk detection (Section~\ref{sub:pedestrianRW}) and guidance (Section~\ref{sub:guidanceRW}).
In both cases, we focus our attention on the technique to convey information to users with VIB. 

In more recent years, commercial applications for orientation and mobility of people with sight impairment became available as well. We briefly describe a few of them in Section~\ref{sub:apps}.
Finally, in Section~\ref{sub:architecture} we describe the architecture of the \zebra{} application.

\subsection{Solutions for pedestrian crossing}
\label{sub:pedestrianRW}
In 2000, Stephen Se proposed the first technique to recognize pedestrian crosswalks with the goal of supporting people with VIB \cite{se}.
The main limitation of this solution is that it fails to recognize a zebra crossing when its pattern is not completely in the camera field of view, or when it is covered by an object (e.g., a car).
Uddin et al. address this problem and propose a solution to improve the effectiveness of the detection algorithm through bipolarity feature check and projective invariant \cite{uddin1,uddin2}.
These first contributions focus on the computer vision algorithm, and do not address the problem of how to interact with the user.

Successively, Ivanchenko et al. illustrate two techniques for detecting pedestrian crosswalks through the camera of a smartphone.
The first technique focuses on zebra crossing and describes an application that produces an audio tone each time a zebra crossing is recognized \cite{ivanchenko}.
An experimental evaluation with two blind test subjects is presented to assess the ability of an individual to determine whether or not there is a crosswalk at a traffic intersection.
The results shows that both test subjects were able to find the zebra crossing in each one of the $15$ trials.
The second technique is aimed at recognizing United States transverse crosswalks (also known as `two stripes' crosswalks) \cite{ivanchenko2}.
In this solution, the recognition algorithm also detects lateral shift of the person with respect to the two-stripes crosswalk.
The presence of the crosswalk is signaled with a short low-pitched tone, followed by a high-pitched tone. If only one single stripe is detected, a low-pitched tone is emitted. If the second stripe is detected later, and the first one is still in the filed of view, a high-pitched tone is emitted. No sound is generated if no stripe is detected.
After detecting the two-stripes crosswalk, the application reproduces a speech message reporting the position of the person (i.e., inside, on the left or on the right of the crosswalk).
An experimental evaluation conducted with two blind test subjects shows that individuals are able to find the crosswalk and are aware of their position with respect to the crosswalk in six cases out of eight trials.

The two solutions proposed by Ivanchenko et al. are extended by Ahmetovic et al. who, focusing on zebra crossings, propose a technique to compute a set of $9$ qualitative relative positions of the user with respect to the crosswalk, each one corresponding to a user action (e.g., go ahead, rotate right, step left, etc.) \cite{dragan}.
These actions are conveyed to the user in the form of speech messages and can guide the person to the best crossing point (i.e., in the middle of the first stripe).
A qualitative experimental evaluation was conducted with five blind test subjects. These were required to complete two tasks. The first one involved the simple detection of a crosswalk positioned in front of the users. The second one involved the detection and location of a crosswalk on the sides, followed by crossing the road.
All test subjects successfully accomplished the first task, while all users except one accomplished the second one.
Two test subjects reported that a sound-based message would convey information more promptly.

In 2014, Ahmetovic et al. proposed the \emph{ZebraRecognizer} algorithm to recognize zebra crossing \cite{AhmetovicICPR2014}.
The algorithm rectifies the ground plane, hence removing the projection distortion of the zebra crossing features. This allows to compute the position of the user with respect to the crosswalk, producing quantitative measures of the frontal, lateral and angular distances.
This technique focuses only on the recognition algorithm, and does not take into account the user interaction problem.
In this contribution we use \emph{ZebraRecognizer} as the reference detection algorithm (see Section~\ref{sub:architecture}).

\subsection{Solutions for guidance of users with VIB}
\label{sub:guidanceRW}
The problem of guiding a user towards and over a zebra crossing can be seen as a special case of the problem of finding and reaching a target destination in a large space.
In both situations, the user has to search for a target destination, align to the target and walk towards it without deviating too much from the right path.

Fiannaca et al. address this more general problem by presenting a \emph{Google Glasses} application to support users with VIB in finding and reaching a doorway in an open space (e.g., a square or a lecture room) \cite{fiannaca}.
A user study with eight blind test subjects evaluates the usability and effectiveness of two audio guiding modes (sonification and speech).
The sonification mode consists of three high-pitch beeps to indicate that a doorway is visible, and three low-pitch beeps if no doorway is in the camera field of view.
In the speech mode, the phrases ``Door found'' or ``No door found'' are reproduced.
In the evaluation, each test subject was asked to reach a doorway about $20$ meters away from a starting point, walking across an open space.
Six tests were conducted both with speech and sonification guiding modes.
Statistically significant results showed that the speech guiding mode leads to a faster discovery ($39.9$\%) and guidance ($34.5$\%).

In our contribution we show different experimental results with no significant differences
between speech and sonification. This can be due to a number of factors, including the type of sonification and the context of application, which will be discussed in the following sections.

In the domain of wearable devices, it is worth examining SWAN (a System for Wearable Audio Navigation) \cite{wilson2007}.
It was designed to assist pedestrian navigation and orientation for people with VIB.
SWAN includes a hardware equipment, positioned in a backpack, that determines user's location and heading direction.
As a result of an extensive evaluation on SWAN's sonification techniques, Tran \cite{tran}, Walker and Lindsay \cite{walker_lindsay}, \cite{wilson2007} determined the characteristics of three non-speech signals. ``Beacon sounds'' are used to reach a desired destination, and are virtually placed at waypoints along a route from the users current location to the selected destination.
``Object sounds'' indicate features in the environment that could potentially be of interest or hazardous.
Finally, ``surface transitions sounds'' denote changes in the surface the user is walking on, and/or important boundaries (e.g., transition from sidewalk to street).
Among other results from the SWAN project, it emerges that non-speech beacons are adequate to present simultaneously different streams of information (e.g., guiding instructions and description of the context).
In our contribution we use this result as a starting point in the design of a part of the sonification that, as we describe in Section~\ref{sec:guidingModes}, simultaneously informs the user about the distance from the target, and the distance from the lateral border of the zebra crossing.

\subsection{Commercial applications to support independent mobility}
\label{sub:apps}
Currently, most commercially available solutions to support orientation and independent mobility are developed in the form of applications for mobile devices (in particular smartphones).
This is due to two main factors. Firstly, over the last five years, mainstream smartphones have become popular among people with VIB thanks to the built-in universal access technology (e.g., Voice Over for iOS and TalkBack for Android).
Secondly, hardware peripherals and software libraries for context management (e.g., reverse geocoding, $k$-NN queries on points of interest) make it relatively easy to develop applications to support orientation and mobility.

According to our experience in the field, the most noteworthy applications in this category are iMove, Ariadne GPS and BlindSquare.
iMove\footnote{At the time of writing, iMove is available for free download from AppStore: \url{https://itunes.apple.com/en/app/imove/id593874954?mt=8}} localizes the position of the user through GPS, and reads the current address, heading and speed.
iMove can also provide a list of points of interest in the surroundings (e.g., shops, schools, bus stops, etc.).
This is useful both to support orientation (e.g., on a known path) and to find out new points of interests while walking or traveling.
A third functionality allows a person with VIB to record speech memos related to a certain position.
The memo is played back whenever the person is in the same place.
This enables tagging of reference points (e.g., an intersection, a bus stop, etc.) that are essential for autonomous mobility.
iMove notifies the user about points of interest and speech memos following a set of preferences related with spatial distance and time. 

Ariadne GPS works similarly to iMove, however it does not impart information about the surrounding points of interest and the speech memos.
Differently from iMove, it provides a map that can be explored by sliding the finger on the touchscreen of the smartphone.
The names of the streets are read by a text-to-speech algorithm.

BlindSquare is analogous to iMove. It does not include speech memos, but it enables users to interact with the Foursquare social network.
It is worth noting that Blindsquare \emph{reads} text messages through its own high quality speech synthesizer.

These three applications use only speech messages, with the single exception of Ariadne GPS, which adopts an alert sound to draw the attention of the user on the upcoming speech message.
We believe that these applications could be much more effective in supporting orientation and guidance if they integrated real time recognition of physical features, like pedestrian crosswalks.
The internal structure of a system to detect these features is described in Section~\ref{sub:architecture}.

\subsection{\zebra{} System Architecture}
\label{sub:architecture}
\begin{figure}[t!]
	\centering
		\includegraphics[width=.4 \columnwidth]{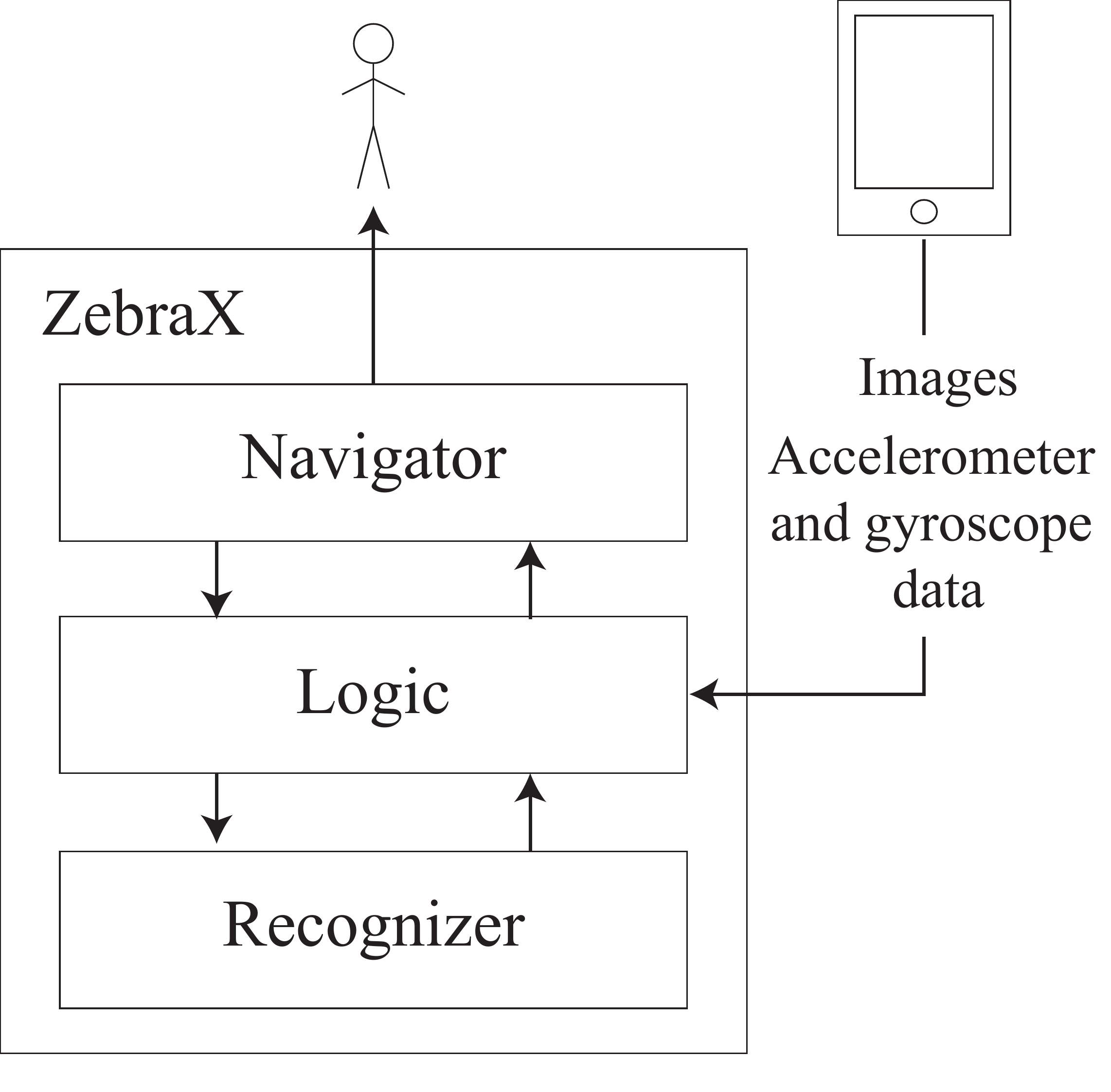}
	\caption[]{\zebra{} architecture}
	\label{fig:architecture}
\end{figure}

\zebra{} is divided into three main modules, as illustrated in Figure~\ref{fig:architecture}.

\begin{figure}[t!]
	\centering
		\includegraphics[width=.5 \columnwidth]{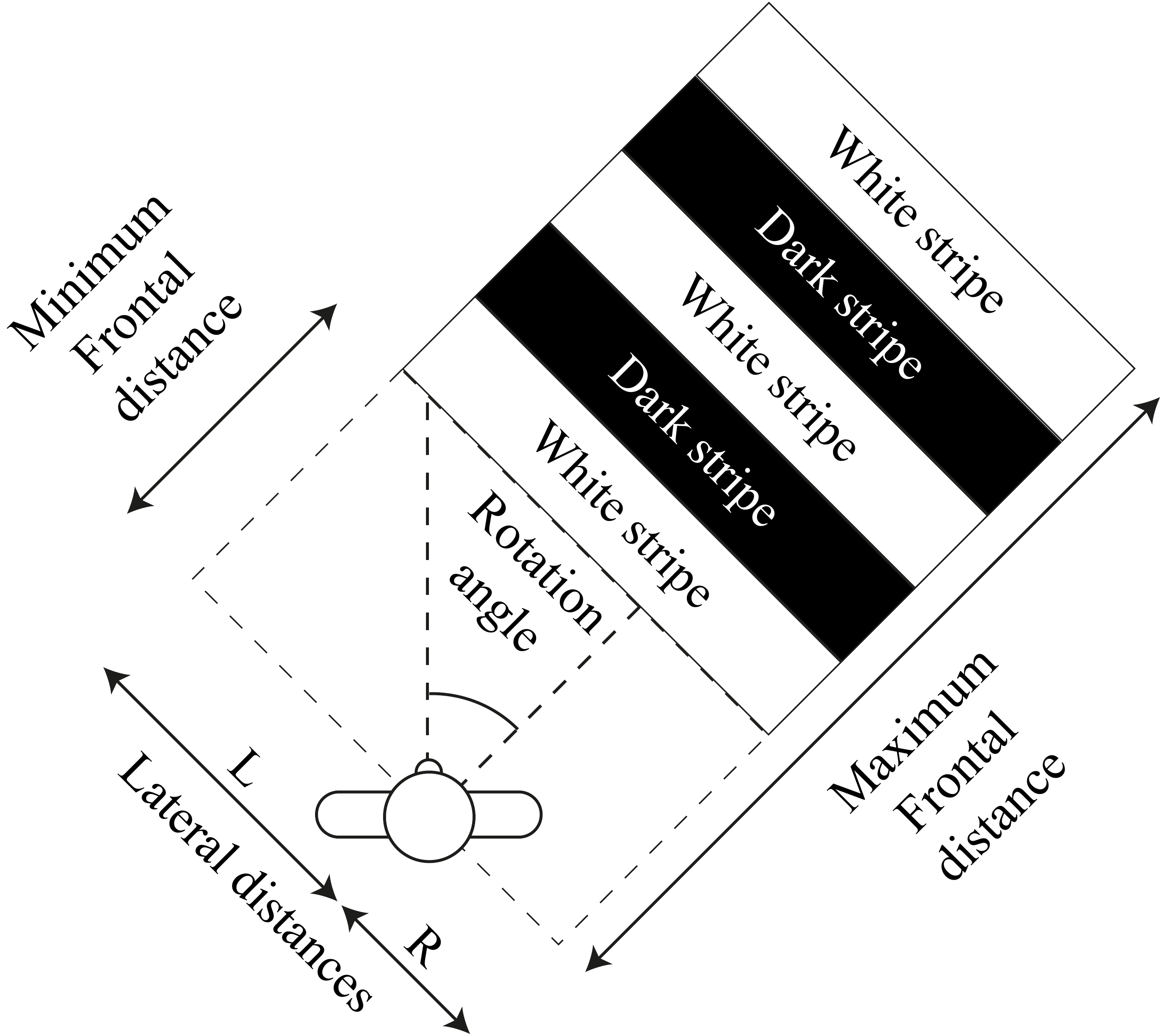}
	\caption[]{Relative distances between user and zebra crossing.}
	\label{fig:distances}
\end{figure}

The \recognizer{} module implements the \emph{ZebraRecognizer} algorithm \cite{AhmetovicICPR2014} which computes the relative distance between the user and the zebra crossing.
In particular, the algorithm computes five measures (see Figure~\ref{fig:distances}).
`Horizontal rotation angle' is the angular distance between the user's heading and the line perpendicular to the stripes.
`Minimum frontal distance' (`maximum frontal distance', respectively) is the distance between the user and the closest (farthest, respectively) stripe.
Finally, `lateral distance left' (`lateral distance right', respectively) is the distance between the user and the left (right, respectively) border of the crosswalk. 

Starting from the positioning data computed by the \recognizer{} module, the \logic{} module computes the messages that are to be conveyed to the user.
There is a total of $7$ messages about the relative position of the crosswalk: `rotate left', `rotate right', `step left', `step right' `not found', `crosswalk ahead', `cross'.
There are two additional messages that help the user to hold the device in the correct position: `raise' and `lower'. These two messages are related to the `vertical rotation angle' (i.e., the device pitch angle), that is computed by the \logic{} module through the accelerometer data.

The \logic{} module is also in charge of keeping distance quantities updated.
Frontal and lateral distances are updated each time the \recognizer{} module completes a recognition cycle (with an average `recognition frequency' of $10$ frames per second).
Vice versa, the value of `horizontal rotation angle' is also updated by using values from the gyroscope. In practice, between two consecutive runs of the recognition algorithm the `horizontal rotation angle' is estimated by correcting the value obtained from the last recognition with the angular distance between the current heading and the last recognition heading.
Intuitively, this solution allows for a rather precise estimate, since the value is reset at every run of the recognizer (i.e., approximately $10$ times a second), and the error introduced by the use of the gyroscope is, for short-time frames, negligible.
The `horizontal rotation angle' is therefore precisely estimated at a frequency (approximately $30$ times a second) that is much higher than the recognition frequency.
Similarly, the `vertical rotation angle' is updated each time the accelerometer data is updated, with a frequency of approximately $30$ times a second.

This contribution focuses on the \navigator{} module, which is in charge of the interaction with the user by acquiring input through the touchscreen, and delivering audio and haptic (vibration) feedback.
\navigator{} has access to the messages computed by the \logic{} module, as well as to the current distance quantities.
Each time any of the distance quantities change, or the current message changes, the \logic{} module notifies the \navigator{} module.


\section{Auditory guiding modes}
\label{sec:guidingModes}
This section presents three auditory guiding modes: speech, mono and stereo.
The first one is based on speech messages, while the second and the third ones are based on mono and stereo sonification, respectively.

Note that this section describes the guiding modes as they were used for the preliminary evaluation (see Section~\ref{sec:intEval}).
After the preliminary evaluation, the changes described in Section~\ref{UpdSon} were applied. The audio files of the final sonifications and examples of their application during road crossing are available on-line\footnote{\url{http://webmind.di.unimi.it/zebraexamples/}}.

\subsection{Speech guiding mode}
\label{SpeechMessages}
Referring to the instructions computed by the \logic{} module (see Section~\ref{sub:architecture}), the \navigator{} module delivers to the user a set of messages generated by the iOS on-board text-to-speech synthesizer.
Since the subjects who participated to the evaluation were all Italian mother-tongue, the messages were delivered in Italian (an English translation is available between brackets).

\begin{itemize}
\item Abbassa/alza il dispositivo (Rise/lower the phone)
\item Ruota a sinistra/destra (Rotate left/right)
\item Passo a sinistra/destra (Step left/right)
\item Non trovato (Crosswalk not found)
\item Strisce davanti (Crosswalk ahead)
\item Attraversa (Cross)
\end{itemize}

Each message is reproduced once, as soon as the \logic{} module computes an instruction different from the previous one.

\subsection{Guiding modes based on sonification}
Sonification is the use of non-speech audio to convey information. A large variety of sonification techniques exist and are used in various applications \cite{kramer1993auditory,csapo}.
The following sections contain an outline of the requirements for the sonifications, followed by a description of the two sonification techniques implemented in \zebra{}.

\subsubsection{Rationale and objectives of the guiding modes based on sonification}
One of the main problems with the speech guiding mode is that it does not convey quantified information about the relative position between the user and the crosswalk. For example, if the user is instructed to rotate right, he/she does not know how much rotation is required in order to be aligned with the crosswalk.
In theory, it could be possible to design a speech guiding mode in which the quantity is reported (e.g., ``rotate right - 20 degrees''). However, this guiding mode would be much more verbose and, most importantly, it would be clearly impractical to update the quantity associated to the message (i.e., the rotation angle in the above example) while the user is moving.

To overcome this problem, the guiding modes based on sonification must inform the user about the \emph{quantity} associated with the instruction.
For this reason we base our technique on \emph{parameter mapping sonification}  \cite{HermanRitter1999}, which is based on the creation of a link between the data to be rendered and the parameters of a synthesizer (or of any other device which generates or plays back sound).

The process of user-centric analysis of the system raised another important requirement that has a direct impact on the sound design. Most people with VIB are not willing to wear headphones, as this prevents the acquisition of audio information from the environment (e.g., an approaching car).
This problem can be partially solved by using bone-conducting headphones\footnote{Bone-conducting headphones do not occlude the ear canal and, therefore, do not impede the perception of the sounds from the surrounding.}.
However, some users declared to find bone-conducting headphones rather uncomfortable, due to the mentally-demanding task to distinguish the sounds produced by the headphones from the environment sound.

Two solutions have therefore been designed: the mono sonification delivers one monaural audio signal, which is suitable to be played by the device speaker. Vice-versa, the stereo sonification employs sound spatialization in order to allow the user to clearly perceive certain sounds as coming from the left or from the right, therefore to convey information using an additional cue. This sonification requires the user to wear a pair of headphones, and employs, for a determined set of messages, a binaural spatialization approach \cite{HammershoiMoller2002}.
Considering the low resolution of bone-conducting headphones in terms of high frequencies (above 10~kHz), and the complexity of the individual-related features of a full Head Related Transfer Function (HRTF) simulation, the stereo technique was not implemented performing a full spatialization. A simpler approach was taken, modifying the differences in level and time of arrival of the sound at the two ears (i.e., Interaural Level Differences - ILD and Interaural Time Differences - ITD).

Two further requirements emerged during sound design:
\begin{itemize}
\item Since for certain types of messages the understanding of the pitch of the sound is essential, the fundamental frequency of the stimulus had to be easily perceived.
\item For a precise spatialization, the sound had to feature a large and dense spectrum.
\end{itemize}

For these reasons, a custom set of impulsive sounds of short duration was designed and implemented.
The test sound was produced by additive synthesis of $5$ to $20$ harmonic or inharmonic partials (depending on the type of message to be sonified), each implemented by an exponentially damped oscillator. Attack times of all partials was set to $1$~ms. The relative amplitude of the partials followed a roll-off of $-3$ to $-6$~dB/octave, whereas decay times differed depending on both the partial and the sonified message type (a similar approach was employed by \cite{KatzRio2008}).
Different repetition and envelope patterns were also used in order to allow a clear distinction between the sonification of the different instructions.

\subsubsection{Mono sonification}
\label{sub:mono}

In order to deliver left-right-type messages without relying on sound spatialization, low pitch sounds were associated to a rotation/step towards the left, and high pitch sounds towards the right. This choice can be intuitively explained considering the keyboard of the piano from the point of view of the player (high-pitch notes on the right).

Considering the list of speech messages in Section ~\ref{SpeechMessages}, the following mono sonifications have been designed and implemented:

\begin{itemize}
\item \emph{Rise/lower the phone}. Impulsive sound with fast transients and harmonic spectrum (similar to a short beep). Two quick repetitions with no pause. High pitch (800~Hz) for the 'rise' message and low pitch (200~Hz) for the 'lower' message. The signal is repeated increasing linearly the rate (from 1~Hz to 2.5~Hz) the closer the user gets to the right inclination.
\item \emph{Rotate left/right}. Impulsive sound with fast transients and in-harmonic spectrum (similar to a percussive sound on metal). The left-right information is delivered modifying the frequency of the stimulus; 300~Hz for the left rotation and 1200~Hz for the right rotation. The repetition rate of the sound is modified linearly from 1.6~Hz (large rotation) to 3.3~Hz (small rotation), varying continuously until the user reaches the target angle.
\item \emph{Step left/right}. Impulsive sound with fast transients and in-harmonic spectrum (similar to a percussive sound on wood). Two fast (200~ms) repetitions. The left-right information is delivered modifying the frequency of the stimulus; 300~Hz for the left step, and 1200~Hz for the right step.
\item \emph{Not found}. Low frequency (200~Hz) in-harmonic sound, slow transients, two repetitions (300~ms the first and 500~ms the second).
\item \emph{Crossing ahead}. Pure-tone (single frequency with no harmonic components) impulsive sound. A rising scale of 6 notes (between 800 and 1700~Hz, one each 100~ms) for a required 10~m advance, 5 notes for 8~m, 4 notes for 6~m, 3 notes for 4~m and 2 notes for 2~m. The scale is repeated every 1000~ms, modifying the message as the person gets closer to the target.
\item \emph{Cross}. Impulsive sound with fast transients and in-harmonic spectrum (similar to a percussive sound on wood). A group of three notes (one note every 150~ms) with fundamentals at 500-800-1000~Hz is repeated every 1200~ms. If the user is required to proceed towards the right, the frequency of the fundamentals is divided by 0.33 (lower pitch), while if towards the right is multiplied by 2 (higher pitch). The level of the sound is rather low, but it becomes louder (up to +20~dB) the more the user needs to modify the path towards the left or the right. When the user is at less than 4 meters from the target, the delay between repetitions is decreased linearly (down to 700~ms).
\end{itemize}

\subsubsection{Stereo sonification}
\label{sub:stereo}

In the stereo sonfication mode the audio signal is delivered differently to the two ears. The user is therefore able to clearly localise a sound in any position between left, center and right. As outlined earlier, the spatialization was performed employing ILD (from 0 to 10~dB) and ITD (from 0 to 0.5~ms)

The following stereo sonifications have been designed and implemented:

\begin{itemize}
\item \emph{Rise/lower the phone}. Same as mono mode.
\item \emph{Rotate left/right}. Same sound as mono mode, frequency 500~Hz. The impulse is continuously repeated every 400~ms, and is spatialized on the left if the user needs to turn left, and vice-versa if the user needs to turn right. The repetition continues until the user can center the sound on the front (therefore when reaching the target angle).
\item \emph{Step left/right}. Same sound as mono mode, frequency 500~Hz. Sound spatialized on the left or on the right (depending on the required direction)
\item \emph{Not found}. Same as mono mode.
\item \emph{Crossing ahead}. Same as mono mode.
\item \emph{Cross}. Same sound as mono mode, with frequencies 500-800-1000~Hz. The left-right direction is given by gradually spatializing the sound on the left or on the right, so that the task of the user is to rotate in order to keep the sound central.

\end{itemize}


\section{Preliminary evaluation}
\label{sec:intEval}
During the design of the auditory guiding modes several test subjects were asked to use the application and provide feedback.
In addition to these informal evaluations, a preliminary evaluation was carried out in order to allow for the fine tuning of the whole application, and in particular of the auditory guiding modes.
This section describes the evaluation methodology, its results and how the guiding modes were changed according to this evaluation.

\subsection{Evaluation methodology}
\label{sub:prelimMethod}
The evaluation was conducted at the Milan Institute for Blind People (Istituto dei Ciechi di Milano\footnote{\url{http://www.istciechimilano.it/}}), which offered support in terms of location and test subjects with VIB for the evaluations.

The evaluation was conducted with five congenitally blind test subjects in a controlled environment, namely a large corridor ($20$m long, $6$m wide approximately), where a real-size zebra crossing was represented on a large plastic sheet.
The choice of conducting the evaluation in an indoor space was driven by the fact that, in this preliminary evaluation, we wanted the test subjects to focus on the sonified audio, without being distracted from environmental noise.
The auditory guidance information was delivered using a pair of wired bone conducting headphones\footnote{Headphones model is \emph{Goldendance Audio Bone Aqua}}, connected with an iPhone $5$.
Each test subject was asked to perform five tasks in random order, one task for each one of the instructions listed in Section~\ref{SpeechMessages} (except for \emph{Not found}).
The goal of each task was to reach a target position (e.g., by rotating, by moving forward, etc.) starting from a random position.
Each task was repeated three times, once for each auditory guiding mode (again, in a random order), and was preceded by a five minutes training.

The following data was measured for each task and auditory modality: time to perform the task, average error (distance from the target, in degrees or metres), and tolerance (number of times each person entered and exited a small area around the target).

At the end of the evaluation, every test subject was asked to give feedback about the application, in particular about the three auditory guiding modes.

\subsection{Results}
\label{PrelRes}

Considering the low number of test subjects, statistical significance was not calculated. Based on simple descriptive statistics, we observed that in the tasks concerning rotation (i.e., rotate left/right and raise/lower the phone) the two sonification guiding modes were more effective than speech guiding mode.
Regarding the other instructions, no notable difference was observed among the three audio guiding modes.

Regarding the test subjects' feedback on the application, it is worth noting that all of them reported to be unable to judge the effectiveness of speech and sonification guiding modes in the real world (i.e., with traffic noise).
To address this problem, successive evaluations (see Section~\ref{sec:finalEval}) were conducted in outdoor space, with audible traffic noise.
Furthermore, the following  comments were made by more than two subjects:

\begin{enumerate}
\item The sound spatialization was not evident. It was often not possible to clearly distinguish when a sound was coming from the left, center or right.
\item The repetition rate changes, which for certain sonified messages indicated the proximity to the target, were not clearly identifiable.
\item Both sonifications required longer training if compared with the speech messages.
\end{enumerate}

In addition to these comments, we observed that in some cases the headphones wire entered the camera field of view, hence preventing the computer vision technique to work properly.

\subsection{Updated auditory guiding modes}
\label{UpdSon}
Certain features and parameters of the auditory guiding modes were modified in order to reflect the results of the preliminary evaluation.

To address the first comment, a simple evaluation was carried out in order to establish the minimum detection thresholds for ILD and ITD using bone conducting headphones.
Using a simple up-down 1~dB step adaptive procedure \cite{Levitt1978}, the discrimination threshold was measured for seven test subjects. The mean discrimination value (i.e., the smallest inter aural difference which allowed a test subject to position a sound source on the left or on the right) for the ILD was 1.15~dB, and for the ITD 0.13~ms.
Considering that these results are sensibly larger to the ones obtainable with a standard pair of headphones, the spatialization ranges were changed. The ILD was increased to a maximum of 20~dB (before it was 10~dB), and the ITD to a maximum of 1~ms (before it was 0.5~ms).

To address the second comment, the following minor modifications have been applied:
\begin{itemize}
\item \emph{Rise or lower the mobile phone} - the repetition rate has been increased to a maximum of 3.3~Hz (before it was 2.5~Hz).
\item \emph{Step left or right} - the repetition rate has been linked to the required displacement (before, the sonification was of boolean type, therefore no information was delivered about the amount of required displacement). The stimulus is repeated every 800~ms if the required displacement is relatively large (2~m), increasing linearly the repetition rate (up to one repetition each 400~ms) for smaller displacements (50~cm).
\end{itemize}

Finally, considering the third comment, an additional functionality was added to \zebra{}. In all auditory guiding modes, the user can tap on the screen of the device to listen the current instruction through a speech message.
In practice, with the speech guiding mode, upon tapping on the screen \zebra{} repeats the last message that was played.
Vice versa, with mono and stereo, upon tapping \zebra{} provides a speech explanation (using the same messages defined for the speech guiding mode) of the instruction being sonified.
The addition of an optional touch-activated speech message within the sonification guiding modes represents a major change in the design of the guiding modes, which is discussed in Section~\ref{sub:discussion}.


\section{Evaluation of auditory guiding modes}
\label{sec:finalEval}
Considering the difficulties in recruiting test subjects with VIB, we decided to carry out the evaluations also on individuals without VIB.
We conducted three sets of empirical evaluations: a quantitative evaluation with $11$ blindfolded sighted test subjects (Section~\ref{sub:quantitative}), a qualitative evaluation with $12$ blind test subjects (Section~\ref{sub:qualitative}) and, finally, a quantitative and qualitative evaluation conducted with $3$ test subjects with VIB (Section~\ref{sub:quantiquali}).
In Section~\ref{sub:discussion} we report a discussion of the empirical results.

The evaluations were conducted with an iPhone $5$s, and all test subjects wore wireless bone-conducting headphones\footnote{Headphones model is \emph{Aftershoks bluez 2}}.

\subsection{Quantitative Evaluation with Sighted Test Subjects}
\label{sub:quantitative}
The quantitative evaluation was conducted with $11$ blindfolded sighted test subjects.
In the following sections the evaluation settings and methodology are described first, followed by the presentation of the results.

\subsubsection{Evaluation Setting and Methodology}
\begin{figure}[t!]
	\centering
		\includegraphics[width=0.5 \columnwidth]{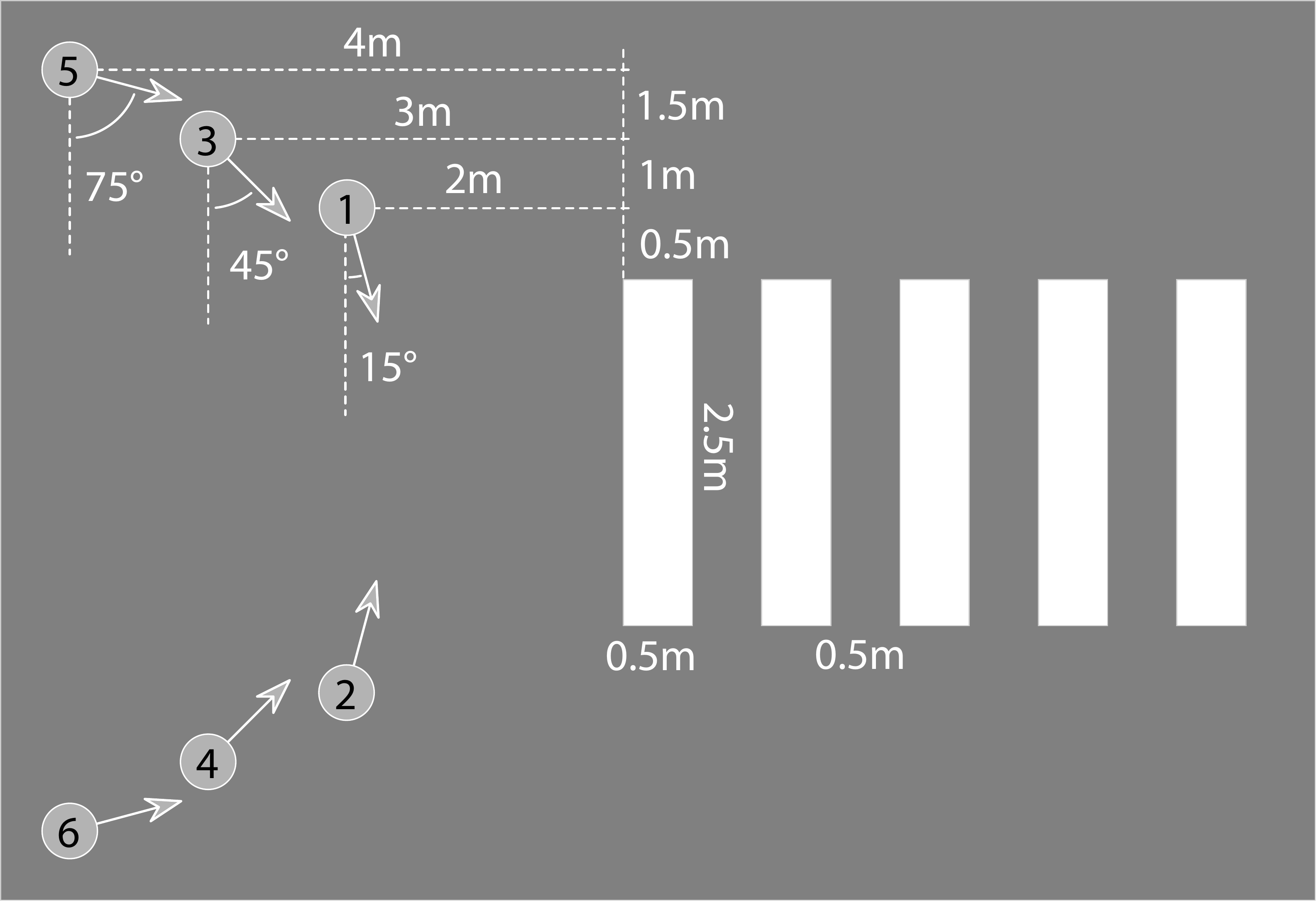}
	\caption[]{Layout of the plastic sheet on which the evaluations were conducted. Numbers and arrows represent starting points and starting directions, respectively.}
	\label{fig:startpts}
\end{figure}

The evaluation was conducted in an outdoor environment where a real-size zebra crossing was represented on a large plastic sheet.
The zebra crossing used during the evaluation is compliant with Italian traffic regulations; it is composed by five light stripes over a dark background, and each stripe is $2.5$m large and $0.5$m wide\footnote{Italian regulation defines zebra crossings that are similar to those used in most countries worldwide} (see Figure~\ref{fig:startpts}).

The outdoor environment was chosen in order to give a more realistic setting to the tests.
In order to reduce the test subjects' ability to orientate using of environmental sounds, and to minimize hazards, it was decided to carry out the evaluation in a large courtyard. Sound of traffic and other environmental noises were audible, but particularly diffuse in the environment, and generally not usable for orientation purposes.
For the same reason, the plastic sheet was moved or rotated after each test, so that it was impossible for the test subjects to predict the position of the zebra crossing based on previous tests.
Furthermore, in order to avoid that tactile and/or audio feedback coming from the ground surface could give clues to help orientation, the whole testing area was covered by a very large plastic sheet

Each evaluation was organized into three phases: learning, practice and measurements.
During the learning phase each test subject had access to a document describing the evaluation structure,
introducing \zebra{} and the three different auditory guiding modes.
The document was presented in the form of an HTML page, so that test subjects could listen to sonification examples\footnote{The document was presented in Italian. Its English translation is available here: \url{http://webmind.di.unimi.it/zebraexplanation/}}.

During the practice phase, each test subject could try \zebra{} with the three auditory guiding modes.
No time constraints were enforced; each test subject could freely decide how long to practice with each guiding mode, until he/she felt comfortable with it. On average, test subjects tested the speech guiding mode for about $1$ minute, and the other two guiding modes for about $2$ minutes each.

During the measurement phase each test subject was asked to autonomously align with the zebra crossing and to actually cross it.
These two operations were repeated for two ``rounds'' of tests. During each round, three tests were conducted, one for each guiding mode, in order: speech, mono, and stereo.
For each test, the subject started from a different point, in a different starting direction.
The choice of the starting points was determined by the idea that the time and effort required to find the crossing, align and cross should be almost the same for all starting points. After some informal evaluations, the $6$ starting points depicted in Figure~\ref{fig:startpts} were chosen.

During the measurement phase, the \zebra{} app recorded a number of parameters related with the completion of the task. These included: the time to align (i.e., to reach the first stripe), the time to cross (i.e., from the first stripe to the end of the crosswalk), the complete list of messages and the number of taps on the screen to repeat/clarify the message.

\subsubsection{Evaluation Results}
\label{sub:quantitativeResults}
During the measurement phase all test subjects were able to successfully complete all crossings. The only exception was the test subject $6$ who, during the test with the mono guiding mode - second round, misinterpreted a ``rotate left'' message and walked straight. Since the subject was going to hit a parked car, the supervisor had to stop the test.

\begin{figure}[t!]
	\centering
		\includegraphics[width=1 \columnwidth]{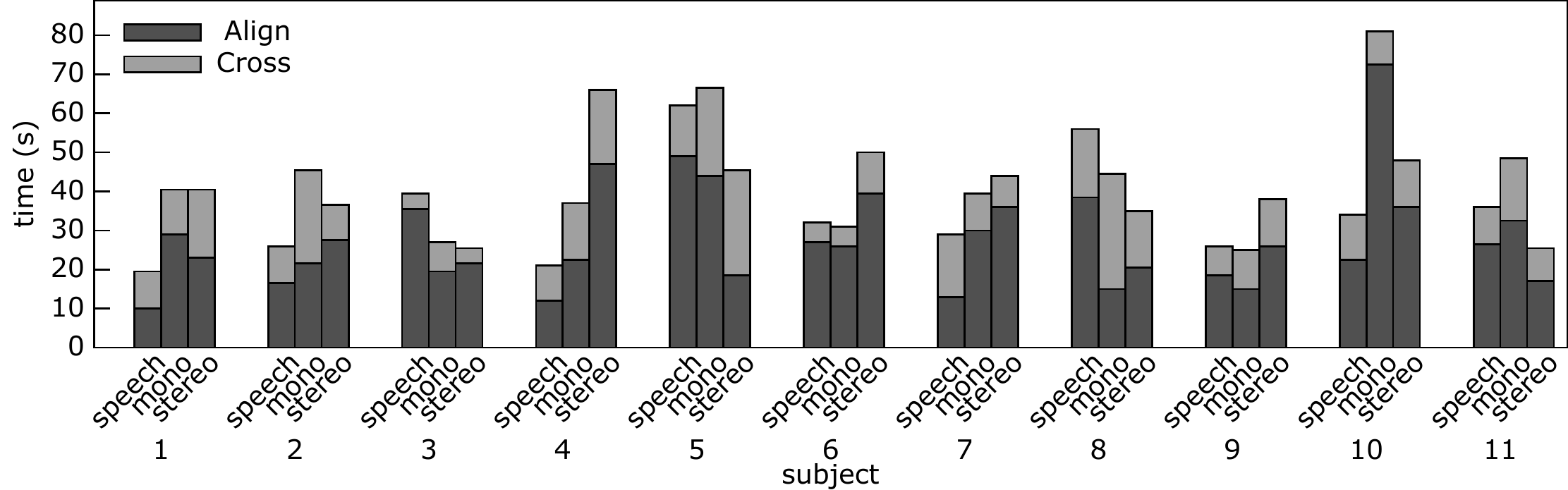}
	\caption[]{Average alignment and crossing time in the two rounds.}
	\label{fig:timeAll}
\end{figure}

Figure~\ref{fig:timeAll} shows, for each test subject and each guiding mode, the average time required in the two rounds to align and cross.
We can observe that $5$ test subjects have been able to align and cross faster with speech guiding mode, $2$ test subjects with mono and $4$ with stereo.
Mean alignment time is $24$s, $29$s and $28$s with speech, mono and stereo modes respectively, while mean crossing time is $10$s, $14$s and $12$s respectively.
Overall, the mean time to align and cross is $34$s, $44$s and $41$s.

\begin{figure}[t!]
        \centering
        \begin{subfigure}[b]{.5\textwidth}
                \includegraphics[width=\textwidth]{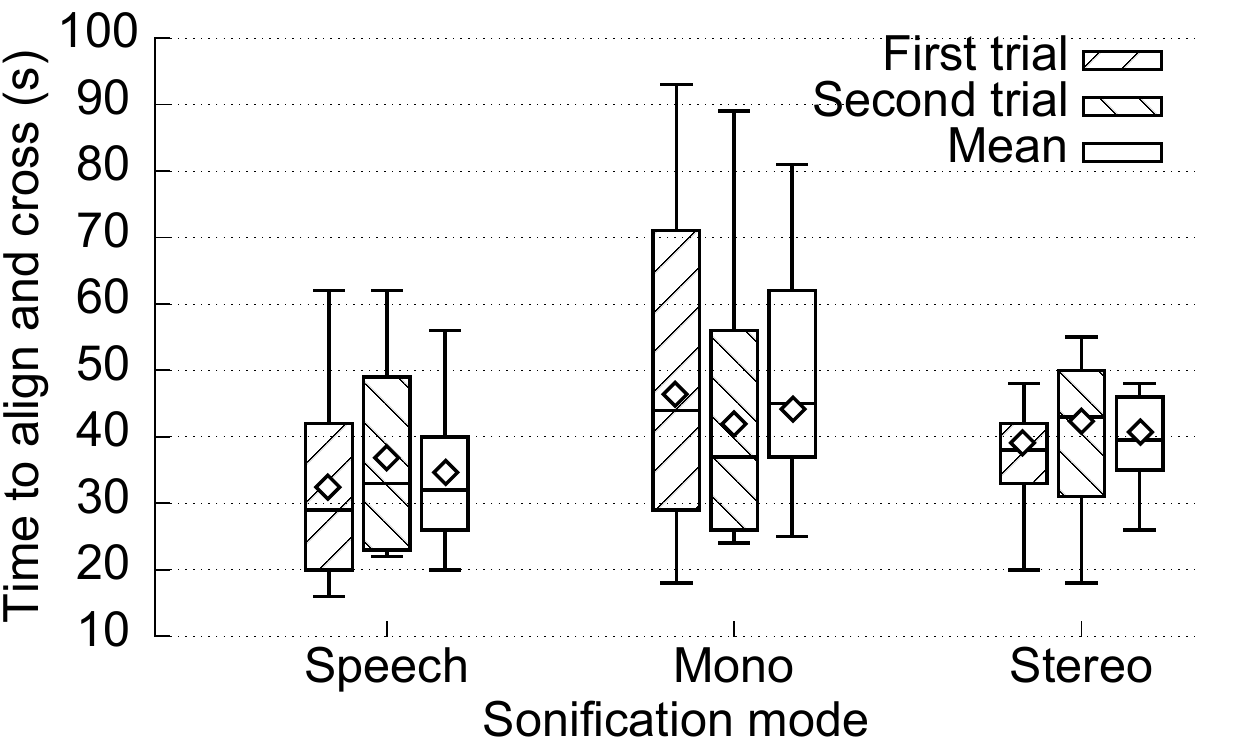}
                \caption{Time to align and cross.}
                \label{fig:timeBoxPlot}
        \end{subfigure}%
        \begin{subfigure}[b]{.5\textwidth}
                \includegraphics[width=\textwidth]{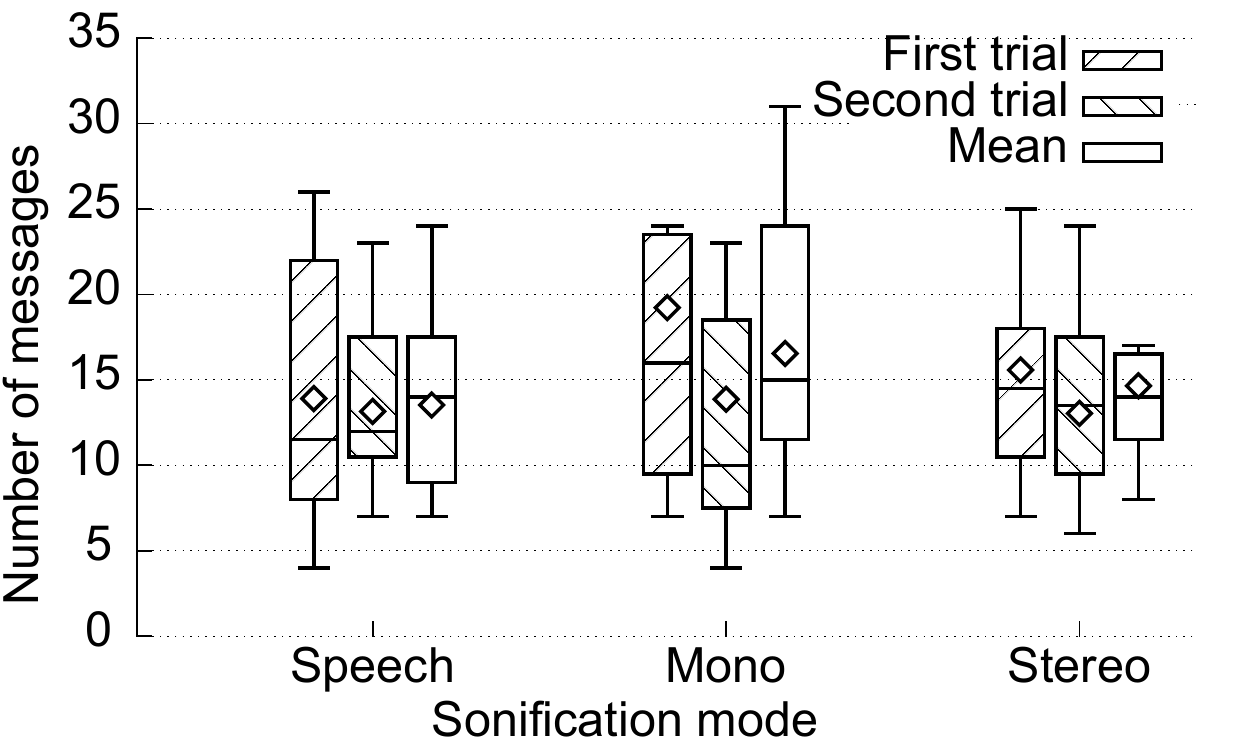}
                \caption{Number of messages.}
                \label{fig:msgBoxPlot}
        \end{subfigure}
        \caption{Boxplot representation ($\diamondsuit$ symbol represents mean).}
        \label{fig:quali}
\end{figure}

The above results seem to suggest that there is not a clear difference in crossing time for the three guiding modes. These results can be also graphically observed in the boxplot shown in Figure~\ref{fig:timeBoxPlot}.
This chart also seems to highlight that, differently from what expected, there is no learning effect between the first and second round. Indeed, on average, the crossing time in the second round is slightly lower for the mono guiding mode compared with the other two guiding modes.

Another metric that can help to understand the performance of the three guiding modes is the total number of changes in the message to be conveyed during the task (this metric will be referred to as ``number of messages''). Clearly, a smaller value indicates higher performance.
In this case it emerges that speech and stereo guiding modes yield very similar results, while mono sonification requires a slightly larger number of instructions, on average (see box plot in Figure~\ref{fig:msgBoxPlot}).

Inferential statistics have been performed to identify whether the differences between guiding mode groups are statistically significant.
Considering the time to align and cross, the data sets are normally distributed, therefore a one-way ANOVA was conducted. The results show that there are no statistically significant differences between the three groups ($F(2,63)=1.178$, $p=0.314$).
Similarly, no statistical difference was found between the first and second round performances, and between the starting points (for all guiding modes).

Considering the number of messages, the data sets are not normally distributed, therefore a Kruskal-Wallis test was conducted. No statistical difference was found between the three groups ($\chi^2 = 0.164, p = 0.921$).

\subsection{Qualitative Evaluation with Blind Subjects}
\label{sub:qualitative}
The qualitative evaluation was conducted in an indoor environment during an exhibition of assistive technologies\footnote{HANDImatica 2014, held in Bologna, Italy.}.
The evaluation was conducted by $12$ blind subjects.

The evaluation was divided into three phases: learning, practice and questionnaire. The learning and practice phases were conducted with the same methodology as the quantitative evaluation.

\begin{figure}[t!]
	\centering
		\includegraphics[width=0.8 \columnwidth]{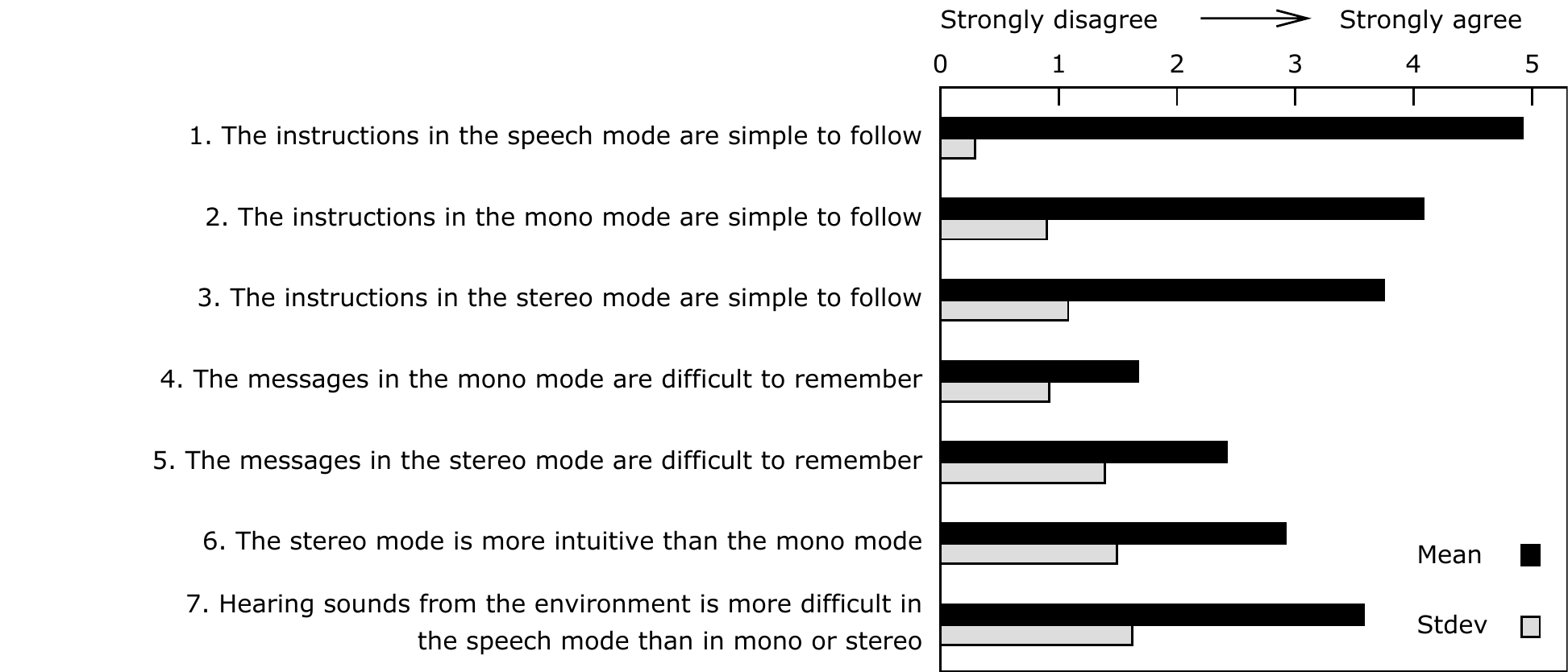}
	\caption[]{Questionnaire and results, first part.}
	\label{fig:quali1}
\end{figure}

The questionnaire is organized in two sets of Likert-scale items; the first one is derived from the System Usability Scale
\footnote{\url{http://www.usability.gov/how-to-and-tools/methods/system-usability-scale.html}},
and is composed of $7$ statements related to the ease of use of the three auditory guiding modes (see  Figure~\ref{fig:quali1}).
The second one is composed of $8$ statements, is derived from IBM Computer Usability Satisfaction Questionnaire (CSUQ) \cite{lewis1995ibm} and is aimed at evaluating the satisfaction with the preferred guiding mode, which is specified by the subjects with an answer to a multiple choice question.

\begin{figure}[t!]
	\centering
		\includegraphics[width=0.8 \columnwidth]{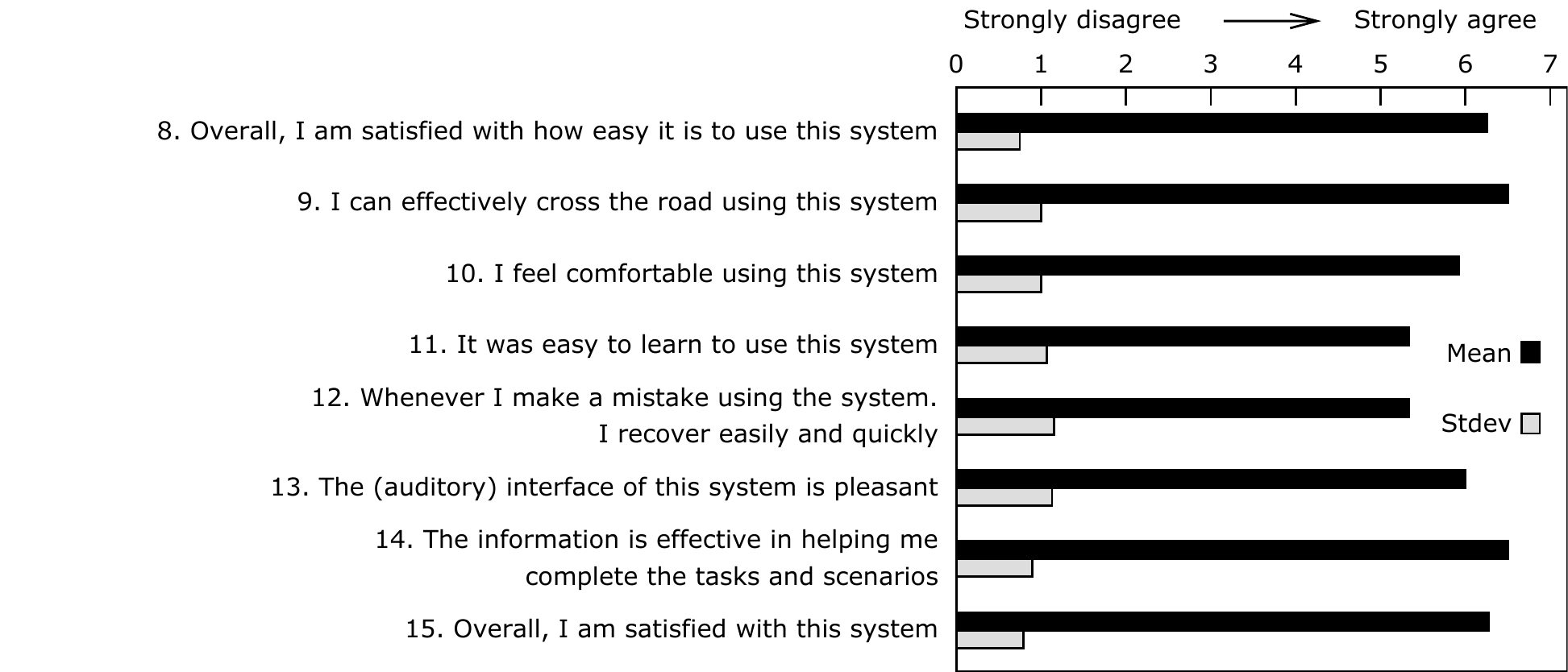}
	\caption[]{Questionnaire and results, second part.}
	\label{fig:quali2}
\end{figure}

There are some topics on which most of the test subjects seem to agree, and others in which there is no consensus.
The test subjects agree on the fact that instructions provided with the speech guiding mode are simple to follow (consider item $1$ in the first set), and they all seem to have an overall positive view of \zebra{} (consider in particular items $1$, $2$, $7$ and $8$ in the second set).

There is generally a lower consensus on the items in the first set. For example, test subjects have very different feelings about the ease of following instructions with the mono guiding mode. $8$ test subjects state that they are easy to follow (with a rate of $4$ or $5$) while $4$ test subjects do not agree with that statement.
Very similar result are obtained for the stereo guiding mode. $8$ test subjects state that instructions provided with the stereo guiding mode are easy to follow.
Interestingly, only one test subject found the instructions provided with both mono and stereo guiding modes hard to follow. Instead, $6$ test subjects found that one of the two guiding modes based on sonification is hard to follow, while the other one is not. This suggests that test subjects have clear and contrasting preferences. To confirm this, $50\%$ of the test subjects state that mono guiding mode is more intuitive than stereo, while $50\%$ state the opposite.

Three test subjects prefers the speech guiding mode, $4$ prefers mono and $5$ prefers stereo.
Despite this, in the second set of items test subjects converge towards a positive view of \zebra{} (see Figure~\ref{fig:quali2}).
Indeed, subjects argue to be satisfied by the ease of use of the application and that they have been able to complete the crossing using \zebra{}.

\subsection{Qualitative and Quantitative Evaluation with Test Subjects with VIB}
\label{sub:quantiquali}
The third evaluation consisted in a quantitative and qualitative evaluation conducted with three test subjects with severe visual impairments.

\subsubsection{Evaluation Methodology}
The evaluation was conducted with three test subjects: one of them was blind, the other two were partially sighted, and not able to recognize zebra crossing through their residual sight\footnote{The two partially sighted subjects were blindfolded during the test.}.

The evaluation consisted in five phases. The first three phases (learning, practice and measurements) were similar to the quantitative evaluation described in Section~\ref{sub:quantitative}.

The fourth phase was conducted in a urban crossroad, and consisted in a set of about $10$ crossing attempts. A supervisor was constantly supporting the test subjects, in the attempt to avoid any hazard. At each crossing attempt the supervisor guided the test subject to the crosswalk vicinity, and then asked him/her to align with the crosswalk. Once aligned, the test subject had to wait for the traffic light to turn green (this information was provided by the supervisor) and was then asked to cross. In case the crossing was not complete before the traffic light turned yellow, the supervisor was instructed to guide the test subject towards the sidewalk. No formal measurements were collected during this phase. The goal was simply to allow the test subjects to use \zebra{} in a real environment.

The fifth phase consisted in the qualitative evaluation described in Section~\ref{sub:qualitative} with an additional set of open questions.

\subsubsection{Evaluation Results}
\begin{figure}[t!]
	\centering
		\includegraphics[width= .4 \columnwidth]{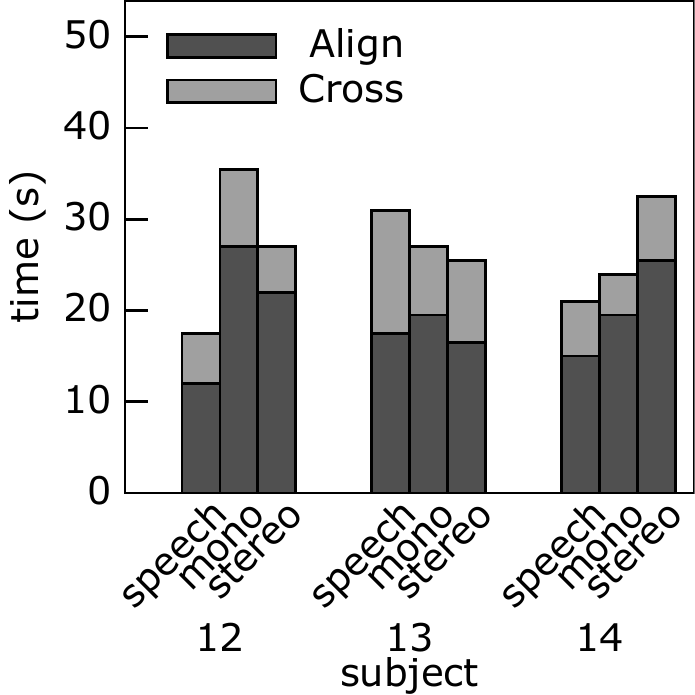}
	\caption[]{Crossing time in the $6$ tests conducted by each of the $3$ test subjects with VIB.}
	\label{fig:timeAll-blind}
\end{figure}

During phase three (measurements), all test subjects have been able to successfully complete the crossing in all the attempts.
Figure~\ref{fig:timeAll-blind} shows the time to align and cross. 
For what concerns the comparison among the three guiding modes, results are not dissimilar to those presented in Figure~\ref{fig:timeAll}.
One difference is that, in the case of test subjects with VIB, the average crossing time is about $27$s with the three guiding modes. This is more than $10$s faster if compared with the performances of blindfolded sighted users.
The number of messages is also similar; mean values are $20$, $11$ and $14$ for the three guiding modes respectively. In this regard, we have to underline that test subject $12$ (the blind subject) had some problems, at the beginning, finding the correct inclination of the device. This caused a large number of `raise' and `lower' messages in the two runs with the speech guiding mode.

In phase four, all test subjects completed the crossing before the traffic light turned yellow. The test subjects conducted at least one test with each guiding mode, but they were left free to choose how to conduct the majority of tests. All of them choose to use their preferred guiding mode (listed below).

In phase five, it emerged that the three test subjects agreed on the fact that the instructions provided in the speech and the mono guiding modes were easy to follow (for both items, two test subjects rated $7$ and the other rated $6$).
A slightly different score was given to the stereo guiding mode (two test subjects rated $4$ and the other rated $3$).
Vice versa, there is no consensus about how hard it is to remember the sonifications; two test subjects reported that they are hard to remember, while test subject $14$ reported the opposite.

Each one of the three test subjects preferred a different guiding mode. Test subject $11$ preferred stereo guiding mode, justifying the choice by saying that the stereo guiding mode ``provides both the spatial references and the clearness of the speech messages that can be activated by tapping''\footnote{The interview was conducted in Italian, and only the english translation is reported.}.
Test subject $12$ declared to prefer the speech guiding mode because it was less cognitive demanding. This test subject comments that ``you need to get used to this app, because when you are crossing you need to pay attention to the surrounding. With the stereo [and mono] guiding mode[s], you need to concentrate to remember the sounds [i.e., the association between the sounds and the instruction], and this may distract you''.
Finally, test subject $13$ preferred the mono guiding mode, reporting these motivations: ``I like the other two [guiding modes] as well. Still, stereo [guiding mode] requires me to concentrate, while speech messages can get confused with other sounds in the environment''.

Finally, the last questions about the overall satisfaction denoted high satisfaction by all three test subjects.

\subsection{Discussion}
\label{sub:discussion}

A number of discussion points emerge from the analysis of the experimental results and from the experience derived by the observation of the different evaluation stages.

It is quite clear that there is no guiding mode which is best suited for all test subjects.
While on average the speech guiding mode allowed the test subjects to align and cross more quickly,
the majority of test subjects ($6$ out of $11$) were faster to align and cross with one or both the sonification guiding modes.
More importantly, test subjects distribute their preferences for the best guiding mode almost uniformly among the three solutions ($4$ prefers speech, $5$ mono and $6$ stereo guiding mode).

An important fact to be considered is that, following the results and feedback of the preliminary evaluation stage (Section~\ref{sec:intEval}), the guiding modes have been integrated with touch-activated speech messages.
While this functionality clearly facilitates the usability of the application, its implementation essentially changed the nature of the evaluation, which in practice became a comparison between a guiding mode based on speech only and two guiding modes based on the combination of sonification and speech.
We expected the test subjects to rely on the tap gesture mainly during the training phase, and then to gradually get used to the sonifications. Nevertheless, we did not observe a statistical significant decrease in the number of tap gestures between the first and second round tasks.

During the tests with the two sonifications, some test subjects frequently tapped on the screen, requesting the speech cue.
We believe that these test subjects did not get well acquainted with the sonification technique, and therefore required constant speech feedback in addition to the sonification.
For example, during the second round with the stereo guiding mode, test subject $4$ tapped on the device almost three times for each new message received ($67$ taps and $24$ messages).
Differently, other test subjects used the tap gesture only sporadically. For example, test subject $5$ tapped only $2$ times in the second round with the mono guiding mode, during which he received $15$ messages in total. This indicates that the test subject was confident to have correctly interpreted the great majority of messages.

Interestingly, sighted test subjects frequently used the tap gesture also with the speech guiding mode (more than half of the sighted test subjects used the tap gesture more than once every four messages).
The tap gesture seemed to provide a form of confirmation or reminder of the last message read. It was not the same for the three test subjects with VIB, who did not use this functionality with the speech guiding mode.

To judge the applicability of the two sonifications we should also consider that, while they are considered less intuitive (all test subjects believed that at least one of the two sonifications was harder to understand than the speech guiding mode), test subjects still expressed their appreciation for them even after a short practice ($11$ out of $15$ test subjects preferred the mono or stereo guiding mode).
This is due to the fact that, according to some of the test subjects, the speech messages prevented the hearing of environment sounds. Also, as reported by two test subjects as answers to the open questions, the guiding modes based on sonification conveyed the ``quantity'' of the expected user movement. This additional information, once appropriately grasped, could further facilitate the alignment and crossing phases.

A further consideration should be made regarding the fact that during the evaluations no learning effect emerged. None of the metrics defined to estimate the time and effort indicated a statistically significant improvement between the two rounds. This could be due to the short duration of the tests. Furthermore, a `tiring' effect could have appeared, considering that the test subjects were required to keep high levels of concentration during the whole evaluation (approximately $20$ minutes). Using the speech guiding mode, $6$ of the $11$ blindfolded test subjects required a longer time to align and cross during the second round if compared with the first one.
Similarly, with both mono and stereo guiding modes $5$ test subjects required longer time in the second run. Since the sonifications appeared to be less immediate, we initially guessed that they should have taken larger benefit from the learning effect derived by frequent use of \zebra{}.

Another aspect to be considered is that the evaluations conducted on the plastic sheet were more challenging than those in the real environment. It is in fact true that when testing the app on the plastic sheet, no specific haptic or audio cue is available. Vice-versa, when crossing on the road there are a number of hints that can help a person with VIB to orientate during the crossing, including, for example, the sidewalk and traffic noise, and the feeling of different types of terrains under the feet.

One final remark is related to the unexpected high dispersion of the quantitative results with respect to the mean. The relative standard deviation is $41\%$, $47\%$ and $40\%$ for speech, mono and stereo guiding modes, respectively.
Combining these data with the experience derived from the observation of the experiments, we can highlight two important facts.
Firstly, some test subjects are more confident and hence move faster (e.g., test subject $9$), while others are more cautious (e.g., test subject $5$) and tend to move and rotate more slowly.
Secondly, there are some human errors that can lead one test subject to have different results in two tests with the same guiding mode.
For example, test subject 4 completed the two tests with stereo guiding mode in $28$s and $104$s respectively. In the second round, the test subject misinterpreted a message, believing that the crosswalk was on his right, while actually it was on his left. This caused the align process to take much longer ($78$s in total) than in the previous round.

\section{Conclusions and future work}
\label{sec:concl}
In this paper we presented two sonification techniques aimed at guiding people with VIB while crossing a road.
The sonifications have been designed with a user-centric approach. The process involved a number of informal evaluations in order to get feedback from the test subjects, as well as a preliminary evaluation, which deeply influenced the design of the sonifications.
The two sonifications have been implemented as part of the \zebra{} prototype, which adopts a state-of-the-art computer vision technique to recognize zebra crossings.
The prototype has then been used to conduct three evaluations on the effectiveness of the two sonifications compared with a less innovative speech-base guiding mode.

Experimental results show that the \zebra{} prototype can effectively guide people with VIB in road crossing with any of the three auditory guiding modes.
Most test subjects ($75\%$) declared to prefer one of the two sonification guiding modes with respect to the speech mode. This result supports the usefulness of the two sonifications, also considering that the test subjects preferred the sonifications despite these being less immediate to use.

At the same time, results show that there is no single guiding mode that is the best for every user.
This was actually expected. In our experience with visual impairments\footnote{One of the authors is congenitally blind, three of the authors are members of a business company developing assistive technologies for people with VIB, and all authors have experience in scientific research on assistive technologies for people with VIB.} we often observed that people with VIB have very different needs, habits and abilities.
This awareness guided the user-centered design phase in which two similar sonifications (mono and stereo) were designed. The main reason for this choice resides in the fact that mono sonification can be also used by individuals who are not willing to wear headphones while walking autonomously.

The results presented in this contribution allow to start working along several directions in research and development.
The three guiding modes find their direct application in commercial software to support road crossing. This software can be a standalone application or a functionality of an application that supports urban orientation, such as, for example, iMove. 

Given the results obtained in this contribution, \zebra{} will allow each user to select his/her preferred guiding mode.
Once the software will available on the market, we can expect that many people will use it\footnote{iMove had more than $100,000$ downloads in two years.}. This will allow to remotely collect usage statistics from a large population. For example it will be possible to monitor how many people use each guiding mode, to collect statistical data about crossing performance, and to know how many people use the application with headphones.
Even more significantly, remote monitoring of real usage will allow to collect data about the long term learning effect, which is a very important aspect, and which we could not practically take into account in this contribution. 
Considering this, further investigations should be carried out about the effects of training on test subjects' performances, in particular using the sonifications. After an adequate training time, the option of the touch-activated speech message with the sonification mode should be eliminated. It is expected that test subjects will not be needing that cue anymore, and will be able to successfully complete the task using sonification-only guidance.
All the information collected remotely will guide the design and fine tuning of future sonifications.

Another possible further development consists in a designing a new guiding mode whose idea derives from the comments of one test subject involved in the test. The novel guiding mode could work as follows. Each new instruction is conveyed through a speech message; additionally, the quantity associated with the message is conveyed to the user through sonification. For example, it would be possible to have a `rotate right' message followed by a sound that informs the users about the quantity of rotation, changing dynamically at the user's rotation. The main difference with the sonifications presented in this contribution is that only one signification is needed for all types of message, simplifying the learning process with the new guiding mode.

The development of sonification techniques will also continue towards different directions, in order to include information other than the sole zebra crossing position. For example, we are currently working on a software module to recognize traffic lights; in the future it will be necessary to convey to the user, at the same time, information about the zebra crossing and about the traffic lights (e.g., their position and the current color - green, yellow, red). With the development of hardware peripherals and of computer vision techniques, it will be possible to recognize more and more aspects of the user's surroundings (e.g., incoming cars, obstacles, etc.), and it will become more and more challenging to design sonification techniques that can effectively transmit this information to the user.

\section{Acknowledgements}
Authors would like to thank the ``Istituto dei ciechi di Milano'' (Milan Institute for Blind People) and all volunteers involved in the empirical evaluation.


\end{document}